\documentstyle[stwol,epsfig]{article}

\def\Journal#1#2#3#4{{#1} {\bf #2}, #3 (#4)}

\def\NPB{{\em Nucl. Phys.} B}

\def\PRL{\em Phys. Rev. Lett.}
\def\PRD{{\em Phys. Rev.} D}

\def\be{\begin{equation}}
\def\ee{\end{equation}}
\def\bea{\begin{eqnarray}}
\def\eea{\end{eqnarray}}

\begin{document}

\title{PROMPT PHOTON PLUS CHARM PRODUCTION AT NEXT-TO-LEADING ORDER IN QCD}

\author{B. BAILEY }
\address{Physics Department, Eckerd College, St. Petersburg, FL 33711, U.S.A.}
\author{ E. L. BERGER and L. E. GORDON }
\address{High Energy Physics Division, Argonne National Laboratory,\\ 
Argonne, IL 60439, U.S.A.}


\twocolumn[\maketitle\abstracts{
The two particle inclusive cross section for the reaction $p 
+\bar{p}\rightarrow \gamma + c + X$ is studied in perturbative quantum
chromodynamics at order $O(\alpha ^2_s)$. Differential distributions 
are provided for various observables, and a comparison is made with
data from the CDF collaboration.}]

\section{Introduction}

Theoretical study of the reaction $p+\bar{p}\rightarrow \gamma + c + X$ is 
interesting for several reasons.  The two particle inclusive cross section 
probes QCD dynamics in more detail than a single particle inclusive cross 
section.  Specification of the cross section in next-to-leading order in QCD 
is correspondingly more challenging, and data will permit a study of various 
correlations including those in the rapidities and transverse momenta of the 
final photon and charm quark.  Second, the data may permit a direct 
measurement of the charm quark density of the proton if it can be shown, as we 
do~\cite{bg,bbg}, that the hard scattering cross section is dominated by the 
contribution from the $cg$ scattering subprocess, where the incident $c$ quark 
is a constituent of one of the initial hadrons.  Third, it is valuable to be 
able to compare data obtained by the CDF Collaboration~\cite{cdf} on prompt 
photon production in association with charm decay products, such as 
$e^\pm$, $\mu^\pm$ and $D^\pm$, with next-to-leading order predictions.
\footnote{Invited paper presented by E. L. Berger at the 28th International 
Conference on High Energy Physics, Warsaw, Poland, July 25 - 31, 1996.  
Argonne report ANL-HEP-CP-96-84.  This work was supported by the US Department 
of Energy, Division of High Energy Physics, Contract no. W-31-109-ENG-38.}

We have published two next-to-leading order calculations in perturbative QCD 
of the reaction $p +\bar{p}\rightarrow \gamma + c + X$ at collider 
energies~\cite{bg,bbg}.  In most of the data of interest, the charm quark 
carries a relatively large value of transverse momentum, $p_T^c \ge$ 10 GeV.  
Since this value is large when compared to the mass of the charm quark, we 
neglect the mass of the quark, and we include hard-scattering subprocesses in
our calculation in which the charm quark is a constituent of the initial 
hadrons.  For values of $p_T^c$ less than 10 GeV or so, it would be necessary 
to retain a finite mass for the quark and to examine a different set of 
subprocesses in which the massive quark is not a constituent in the 
initial state but is, instead, produced in the hard scattering~\cite{sally}.  A 
calculation with $m_c \ne 0$ also requires resummation of logarithmic terms of 
the form $\alpha_s^n {\rm ln}^n(p_T^c/m_c)$ for large values of $p_T$.  

\begin{figure}
\center
\epsfig{figure=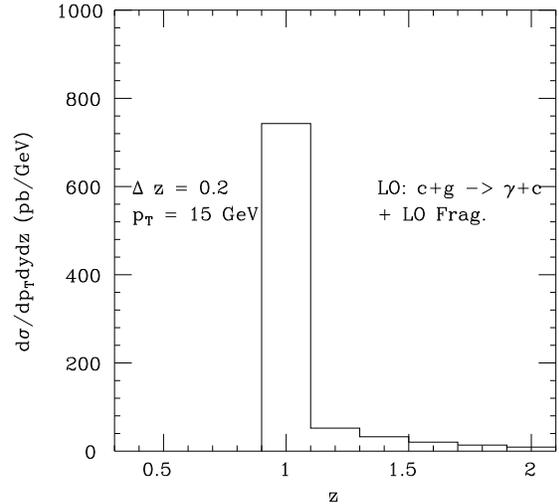,height=2.75in}
\caption{Cross section as a function of z at $y^\gamma=0$ in leading order.}
\end{figure}
In leading-order, the contributions to associated production include the 
direct term $c + g \rightarrow \gamma + c$ along with several photon 
fragmentation processes typified by the hard scattering subprocess 
$c + q \rightarrow c + q$ followed by long-distance fragmentation of the 
final $q$ into a photon: $q \rightarrow \gamma + X$.  In next-to-leading 
order, there are gluon-exchange loop corrections to the lowest-order 
direct process, real gluon emission radiative corrections to the lowest-order 
direct process, and several new  
three-body final-state contributions:  $g + g \rightarrow c + \bar c + \gamma$,
$q + \bar q \rightarrow c + \bar c + \gamma$, 
$c + q \rightarrow c + q + \gamma$, 
$c + \bar q \rightarrow c + \bar q + \gamma$,
$c + \bar c \rightarrow c + \bar c + \gamma$, and 
$c + c \rightarrow c + c + \gamma$.  We include all of these.  In 
principle, we should also include fragmentation to next-to-leading order.  
However, in view of the small size of the fragmentation contribution and the 
theoretical problems identified~\cite{bqiu} with isolated prompt photon 
production in next-to-leading order, we treat the fragmentation terms at 
lowest order.  

The distinction between our two calculations~\cite{bg,bbg} resides in the 
different techniques used in performing the phase-space 
integrals.  In the first paper, we insist on going as far as we can using 
purely analytical techniques.  In the second, we use a combination of 
analytical and Monte Carlo techniques~\cite{bob}.  The second approach  
allows implementation of isolation cuts and other experimentally relevant 
selections.  In both cases, we evaluate the gluon exchange and the real 
gluon emission diagrams analytically, 
expose the infra-red and collinear singularities, 
cancel the infra-red singularities, and factor the collinear singularities into 
incident parton densities or final-state fragmentation functions.  We use 
dimensional regularization to deal with the singularities.  Although 
somewhat restrictive, the purely analytic calculation~\cite{bg} provides the 
size of the two-particle inclusive cross section, the relative importance of 
various contributing subprocesses, and an important check on the more versatile 
combination~\cite{bbg} of analytic and Monte Carlo methods.  

Some of our analysis is done in terms of the variable 
$z= - {\vec p}\,^c_T \cdot {\vec{p}\,^\gamma_T}/(p_T^\gamma)^2$, and some of 
the results are presented as histograms in $z$ for finite intervals in $z$.  If 
$z$ is positive, the photon and charm quark are in opposite hemispheres. If 
$z = 1$, the photon and charm quark have equal and opposite values of 
transverse momentum.  This situation occurs for the leading-order direct 
process, for gluon exchange loop corrections to the leading-order direct 
process, whenever an emitted gluon becomes soft, and when a third parton in a 
three-parton final-state process becomes collinear to an incident parton.  The 
point $z = 1$ is associated with various soft and collinear poles, exposed 
when the cross section is expanded as a function of $z$.  The singularities 
are expressed in terms of delta functions, $\delta(1-z)$, and 
``plus"-distributions, $1/(1-z)_+$.  The singularities are integrable, and 
results are expressed as integrals over finite intervals in $z$.  
\begin{figure}
\center
\epsfig{figure=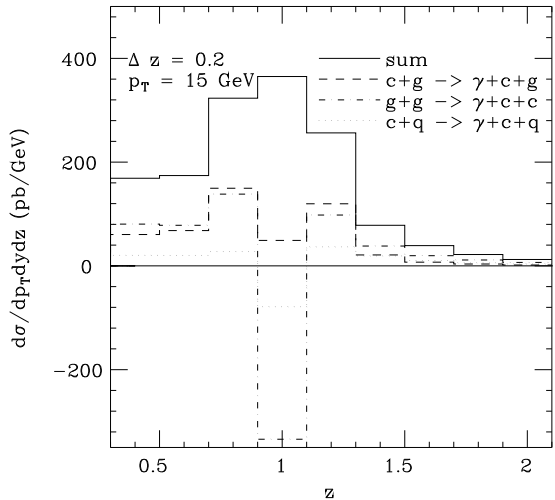,height=2.75in}
\caption{Cross section as a function of z in next-to-leading order.  The 
dominant contributions are shown.}
\end{figure}

\section{Numerical Results}

Our results are presented at a center-of-mass energy $\sqrt{s}=1.8$ TeV.
The renormalization and factorization scales are taken as $\mu=p_T^\gamma$. 
We sum over charm and anticharm production throughout. 
In Fig.~1 the net lowest-order contribution is shown as a function of $z$
at $p_T^\gamma=15$ GeV. The lowest
order cross section is made up of the lowest order direct term
$cg\rightarrow\gamma c$, proportional to $\delta(1-z)$, and fragmentation 
contributions.  In Fig.~1, the direct term provides the peak at $z=1$, and 
the various photon fragmentation terms contribute in the region $z\geq 1$.  
Note the absence of any contribution in the region $z < 1$, an 
unrealistic prediction that illustrates the inadequacy of the lowest order
approach.  The cross section 
in $z$ resulting from our the next-to-leading order 
massless-charm calculation is displayed in Fig.~2.  The next-to-leading order 
collinear initial-state contributions reduce the peak at $z=1$, and 
the next-to-leading order three-parton final-state contributions broaden the 
distribution. The cross section is finite  at all values of $z$, similar to the
situation observed in experiments. In Fig.~2 we also present the individual 
contributions from the most important subprocesses. 
The $cg$ initiated process dominates the cross section, but there are important
contributions from the $gg$ and $cq$ initiated subprocesses in the 
region of small $p_T^\gamma$.  The predicted distribution in $p_T^\gamma$ is 
shown in Fig.~3.  As might be expected, the contribution from the $\bar{q}q$ 
initial state begins to become important at large $p_T^\gamma$.  

Distributions in $p_T^c$ and in the rapidities of the $\gamma$ and $c$ quark 
may be found in our publications~\cite{bg,bbg}.  In the second paper, we 
also discuss the influence of photon isolation, and we compare the magnitude 
and shape of the next-to-leading order and leading order cross sections for 
various observables.  As an example, we present Fig.~4, a distribution in 
$p_T^c$ for different selections on $z$ and the isolation cone variable $R$.  

\begin{figure}
\center
\epsfig{figure=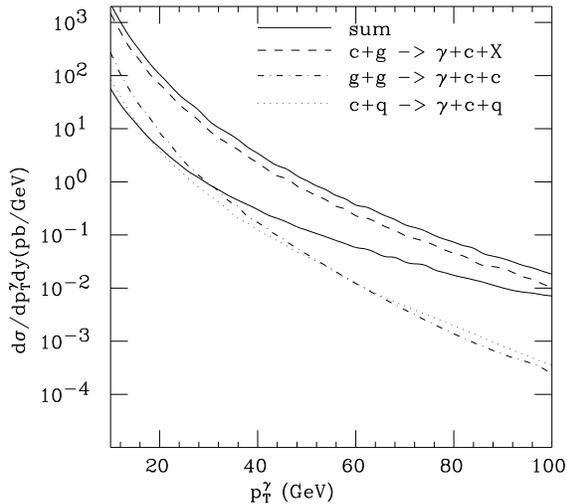,height=2.75in}
\caption{The transverse momentum dependence of 
$d\sigma/dp^\gamma_{T} dy^\gamma dz$, for $z$ integrated over the 
interval $0.2 <z <2.0$.  The upper solid line shows the sum of all subprocesses
through next-to-leading order.  The dashed line shows the sum of the 
$O(\alpha_s)$ and $O(\alpha ^2_s)$ contributions from the $c g$ initial state.
The $O(\alpha ^2_s)$ contributions from the $g g$ and $c q$ initial states are
shown as dash-dot and dotted curves.  The lower solid line shows the 
$O(\alpha ^2_s)$ contribution from the $\bar{q}q$, $\bar{c}c$, and
$c c$ initial states.}
\end{figure}
Positive correlations in rapidity at collider energies are a dynamical 
property of the hard-scattering matrix element~\cite{elb}.  Indeed, it may 
be shown~\cite{elb} that the parton-level matrix element is proportional to 
$ {\rm cosh}^{-1}(\Delta y)$, where ${\Delta y}$ is the difference in the 
rapidities of two final-state particles in a two-particle inclusive reaction.   
In Fig.~5, we present the distribution in the difference of the rapidities of 
the photon and charm quark.  In this example, the $\gamma$ is selected to have 
rapidity in the forward region 1 $< y^{\gamma} <$ 2.  Under these 
circumstances, as may be observed in Fig.~5, the charm quark is predicted to 
have positive rapidity also, with $\Delta y = y^{\gamma} - y^c$ peaking a 
value only slightly larger than 0.   
    
In Fig.~6 we compare our results to CDF data~\cite{steve} for photon plus 
$\mu^\pm$ production, where the $\mu^\pm$ is identified as the decay product 
of a charm quark.  The three upper points are obtained~\cite{steve} from 
the Monte Carlo event-generator Pythia whereas the lower point is our 
theoretical expectation. The Pythia cross section lies substantially below 
the data while our cross section exceeds the data somewhat.  
\begin{figure}
\center
\epsfig{figure=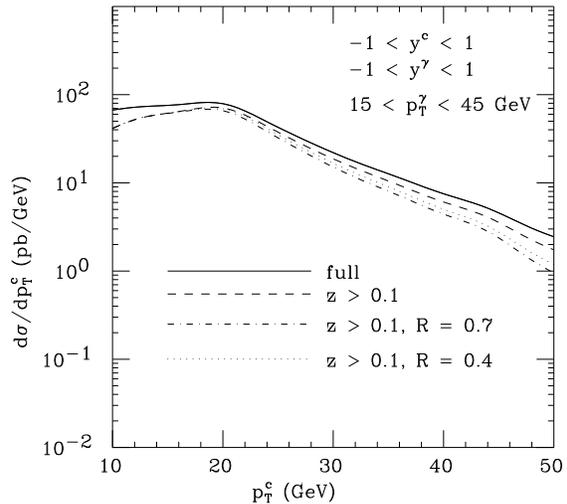,height=2.75in}
\caption{Cross section $d\sigma/dp^c_{T}$ as a function of $p^c_T$, 
with $15 < p^\gamma_T < 45$ GeV, and the rapidities of the photon and charm 
quark restricted to $-1.0 < y < 1.0$.  The solid curve shows the cross section 
with no further restrictions and the dashed curve the result after 
the additional selection $z > 0.1$.  The dotted and dash-dot
curves display the results after photon isolation restrictions are applied, in
addition to the cut on $z$.}
\end{figure}

\section{Conclusions}
We presented the results of next-to-leading order QCD calculations of 
the inclusive production of a prompt photon in association with a charm 
quark.  The two-particle inclusive cross section allows more refined tests of 
the underlying QCD dynamics, revealed, e.g., through rapidity correlations.  
Our cross section shows the expected reduced dependence on the choice of the 
factorization/renormalization/fragmentation scale in QCD.  The sensitivity of 
the cross section to the charm quark content of the proton is preserved in 
next-to-leading order.  A comparison with the CDF data shows reasonable 
agreement.  Further work is underway to include effects associated with the 
finite mass of the charm quark and with charm fragmentation: 
$c \rightarrow D X$ or $c \rightarrow \mu X$.  In other work, we are 
investigating 
associated production of photons and charm quarks at HERA energies and the  
associated production of photons and bottom quarks at collider energies.  

\begin{figure}
\center
\epsfig{figure=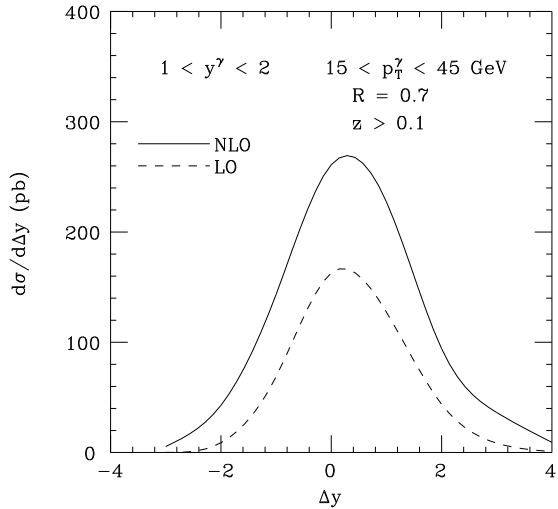,height=2.75in} 
\caption{Cross section $d\sigma/d\Delta y$ as a function of the difference 
$\Delta y = y^\gamma - y^c$ of the rapidities of the photon and charm quark,
for $p +\bar{p}\rightarrow \gamma + c + X$ at $\sqrt{s}=1.8$ TeV.  The
ratio $z$ is restricted to $z > 0.1$, and photon isolation is imposed, with 
$R$ = 0.7.  The transverse momentum and rapidity of the photon are selected 
to be in the intervals $15 < p^\gamma_T < 45$ GeV and $1.0 < y^\gamma < 2.0$.  
The dashed curve shows the behavior at leading order.}  
\end{figure}
%
\begin{figure}
\center
\epsfig{figure=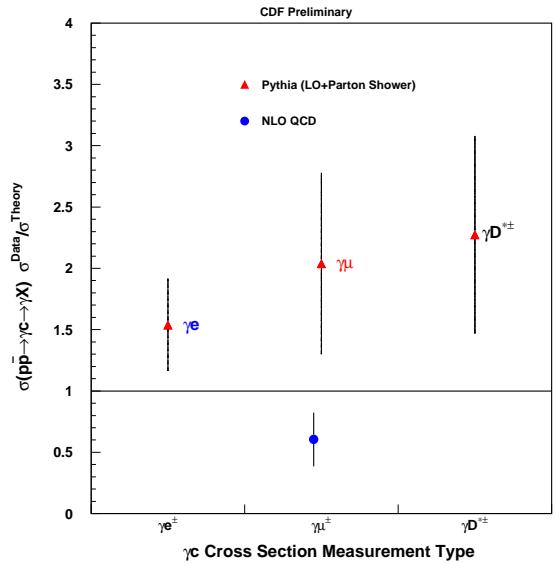,height=3.3in} 
\caption{Ratio of the measured $\gamma c X$ cross section to that predicted, 
for various charm decay channels.}
\end{figure}

\end{document}